\documentclass[pra,aps,twocolumn,amsmath,amssymb,showkeys]{revtex4}
\usepackage{graphicx}
\newcommand{\hh}{{\mathcal{H}}}

\newcommand{\pen}{\openone}
\newcommand{\bro}{{\boldsymbol{\rho}}}
\newcommand{\vbro}{{\boldsymbol{\varrho}}}
\newcommand{\cbr}{{\mathbb{R}}}
\newcommand{\xdif}{{\mathrm{d}}}
\newcommand{\bxx}{{\mathsf{x}}}
\newcommand{\bpx}{{\mathsf{p}}}
\newcommand{\brx}{{\mathsf{r}}}
\newcommand{\bsx}{{\mathsf{s}}}

\newcommand{\mrf}{{\mathsf{R}}}
\newcommand{\msf}{{\mathsf{S}}}
\newcommand{\iu}{{\mathtt{i}}}
\newcommand{\veps}{\epsilon}
\newcommand{\varep}{\varepsilon}
\newcommand{\tiv}{\tilde{v}}
\newcommand{\clom}{\varOmega}
\newcommand{\cln}{{\mathcal{N}}}
\newcommand{\clk}{{\mathcal{K}}}
\begin{document}
\clearpage
\preprint{}

\title{R\'{e}nyi formulation of entanglement criteria for continuous variables}

\author{Alexey E. Rastegin}
\affiliation{Department of Theoretical Physics, Irkutsk State University,
Gagarin Bv. 20, Irkutsk 664003, Russia}

\begin{abstract}
Entanglement criteria for an $n$-partite quantum system with
continuous variables are formulated in terms of R\'{e}nyi
entropies. R\'{e}nyi entropies are widely used as a good
information measure due to many nice properties. Derived
entanglement criteria are based on several mathematical results
such as the Hausdorff--Young inequality, Young's inequality for
convolution and its converse. From the historical viewpoint, the
formulations of these results with sharp constants were obtained
comparatively recently. Using the position and momentum
observables of subsystems, one can build two total-system
measurements with the following property. For product states, the
final density in each global measurement appears as a convolution
of $n$ local densities. Hence, restrictions in terms of two
R\'{e}nyi entropies with constrained entropic indices are
formulated for $n$-separable states of an $n$-partite quantum
system with continuous variables. Experimental results are
typically sampled into bins between prescribed discrete points.
For these aims, we give appropriate reformulations of the derived
entanglement criteria.
\end{abstract}

\keywords{convolution, Fourier transform, R\'{e}nyi entropy, uncertainty principle, Young inequality}

\maketitle

\pagenumbering{arabic}
\setcounter{page}{1}

\section{Introduction}\label{sec1}

Quantum entanglement is one of the fundamental properties of
Nature at the microscopic level. This quantum-mechanical feature
was concerned by founders in the Schr\"{o}dinger ``cat paradox''
paper \cite{cat35} and in the Einstein--Podolsky--Rosen paper
\cite{epr35}. In view of the role of entanglement in quantum
theory, related questions deserve to be studied in detail (see,
e.g., the review \cite{hhhh09} and references therein). Due to
progress in quantum information processing, both the detection and
quantification of entanglement are very important. In the case of
discrete variables, the positive partial transpose (PPT) criterion
\cite{peres96} and the reduction criterion \cite{mhph1999} are
very powerful. On the other hand, no universal criteria are known
even for discrete variables. Say, the PPT criterion is necessary
and sufficient for $2\times2$ and $2\times3$ systems, but ceases
to be so in higher dimensions \cite{horodecki96}. Separability
conditions can be derived from various uncertainty relations
\cite{guhne04,giovan2004,devic05,guhne06,devic07,huang2010,rastsep}.
These studies concerned finite-dimensional quantum systems. For
systems with continuous variables, detection of entanglement is a
more challenging task. Reasons for studying quantum information
with continuous variables are originated in the fact that many
quantum protocols can be efficiently implemented within current
technologies of quantum optics \cite{braunv05,arl14}.

Due to a practical importance, Gaussian states were well studied
from the viewpoint of entanglement detection
\cite{duan00,simon00}. Properties of Gaussian entanglement with
respect to information processing were discussed in
\cite{adesso07,agp10,adesso11,mista15}. Entanglement criteria of
the second-order type deal with variations of certain observables.
For systems with continuous variables, such criteria were proposed
in \cite{duan00,tan99,reid01,mgvt02,gmvt03}. The authors of
\cite{vogel05} formulated an infinite hierarchy of conditions for
positive partial transpose involving higher-order moments. The
conditions of \cite{vogel05} were later amended in \cite{piani06}.
Such conditions provide a very powerful criterion, which is rather
hard for implementation in experimental practice. The authors of
\cite{wtstd09,stw11} have formulated biseparability conditions in
terms of differential entropies related to measurement statistics.
This method differs from some previous studies, in which
entropies of density operators were considered
\cite{barnett89,hhh96,cerf97,wolf02}. In particular, inequalities
with R\'{e}nyi entropies of density matrices can be treated as a
condition for local realism \cite{hhh96}. The authors of
\cite{wolf02} studied the relation between entanglement properties
and conditional R\'{e}nyi and Tsallis entropies for bipartite
quantum systems in finite dimensions. Many covariance-matrix-based
criteria and Shannon-entropy criteria can be unified within a
general formalism proposed in \cite{huang2013}.

When more than two parties are involved, the structure of
entanglement is much richer in comparison with the bipartite case
\cite{gt2009,sv13}. The authors of \cite{sv13} developed a general
framework for constructing multipartite entanglement tests.
Separability eigenvalue equations of \cite{sv13} allow one to witness
partial and full entanglement in multipartite composed systems.
This basic techniques has been applied to examine multipartite
entanglement of frequency-comb Gaussian states \cite{gsv15,gsv16}.
It was shown in \cite{gsv16} that there are two-separable states
which include all other forms of higher-order entanglement. This
example illustrates significance of studying different categories
of multipartite entanglement. In practice, we would often like to
detect the entanglement of states that are partially or completely
unknown. In this case, desired entanglement criteria should
immediately be related to results of some measurements specially
built for such purposes. Another approach is to construct the
density operator via quantum tomography, but quantum tomography
usually requires considerable effort. The separability problem
with partial information was addressed in \cite{huang2013,sped12}.
The formalism of \cite{huang2013} allows one to extend biseparability
conditions to multimode case in both discrete- and
continuous-variable systems.

The aim of this work is to formulate $n$-separability
conditions for an $n$-partite system with continuous variables in
terms of generalized entropies. We also study derived entanglement
criteria from the viewpoint of sampling density functions into
bins. The paper is organized as follows. In Sect. \ref{sec2}, the
required material is presented. Entropic functionals of the
R\'{e}nyi type are briefly discussed. Further, we recall Young's
inequality and its converse, both formulations with sharp
constants. In Sect. \ref{sec3}, we formulate R\'{e}nyi-entropy
entanglement criteria for a multipartite quantum system with
continuous variables. The global observables are constructed
within a commonly accepted approach to deriving separability
conditions for such systems. In Sect. \ref{sec4}, we examine the
presented separability conditions from the viewpoint of their use
in practice of quantum information processing. Appropriate
reformulations are given for sampling measurement statistics into
prescribed bins. In this case, separability conditions in terms of
Tsallis entropies are also given. A utility of the derived
criteria is illustrated with examples in Sect. \ref{sec5}. It is shown
that separability conditions in terms of generalized entropies
sometimes lead to more robust detection of entanglement.

\section{Preliminaries}\label{sec2}

In this section, we recall the required material and describe the
notation. Let $x\in\cbr$, and let $v(x)$ be probability density
function of some continuous variable. Then, the differential
Shannon entropy is defined as \cite{CT91}
\begin{equation}
H_{1}(v):=-\int_{\cbr} v(x)\,\ln{v}(x)\,\xdif{x}
\, . \label{denw}
\end{equation}
There are several fruitful generalizations of the standard Shannon
entropy. For $0<\alpha\neq1$, the differential R\'{e}nyi entropy
is written as
\begin{equation}
H_{\alpha}(v):=\frac{1}{1-\alpha}
\>\ln\!\left(
\int_{\cbr} v(x)^{\alpha}\,\xdif{x}
\right)
 . \label{redenw}
\end{equation}
This entropy is a continuous analog of the $\alpha$-entropy
introduced by R\'{e}nyi in \cite{renyi61}. Entropies of
discrete random variable will be used, when continuous variable is
sampled into chosen bins. Let such bins be specified by the set of
marks $\{\xi_{i}\}$. Hence, we have the intervals
$\Delta\xi_{i}=\xi_{i+1}-\xi_{i}$ with the maximum
$\Delta\xi=\max\Delta\xi_{i}$. We then introduce probabilities
\begin{equation}
q_{i}:=\int\nolimits_{\xi_{i}}^{\xi_{i+1}}
v(x)\,\xdif{x}
\, . \label{qivx}
\end{equation}
To get a good exposition, the size of bins should be sufficiently
small in comparison with a scale of considerable changes of
$v(x)$. For the discrete distribution with probabilities
(\ref{qivx}), its R\'{e}nyi $\alpha$-entropy is defined as
\cite{renyi61}
\begin{equation}
H_{\alpha}(q)
:=\frac{1}{1-\alpha}\>\ln\!\left(\sum\nolimits_{i} q_{i}^{\alpha}\right)
 , \label{rpdf}
\end{equation}
where $0<\alpha\neq1$. In the limit $\alpha\to1$, we obtain the
usual Shannon entropy
\begin{equation}
H_{1}(q):=-\sum\nolimits_{i} q_{i}\,\ln{q}_{i}
\, , \label{spdf}
\end{equation}
where $-\,0\ln0\equiv0$ by definition. Many
interesting properties of R\'{e}nyi entropies with applications
are discussed in \cite{bengtsson}.

To pose required mathematics formally, we introduce convenient
norm-like functionals. For arbitrary $\alpha>0$, one defines
\begin{equation}
\|f\|_{\alpha}:=
\left(
\int_{\cbr} |f(x)|^{\alpha}\,\xdif{x}
\right)^{1/\alpha}
 . \label{alpnrm}
\end{equation}
Of course, we will further assume that such integrals exist. The
right-hand side of (\ref{alpnrm}) gives a legitimate norm only for
$\alpha\geq1$. The case $\alpha=\infty$ is allowed and leads to
the essential supremum \cite{lieb01}. For the given
discrete distribution and $\alpha>0$, we also define
\begin{equation}
\|q\|_{\alpha}:=\left(
\sum\nolimits_{i} q_{i}^{\alpha}
\right)^{1/\alpha}
 , \label{qanrm}
\end{equation}
including $\|q\|_{\infty}=\max{q}_{i}$. The $\alpha$-entropy
(\ref{redenw}) can be rewritten as
\begin{equation}
H_{\alpha}(v)=\frac{\alpha}{1-\alpha}
\>\ln\|v\|_{\alpha}
\, . \label{redenw1}
\end{equation}
In a similar manner, we express (\ref{rpdf}) in terms of
(\ref{qanrm}). Due to $q_{i}\leq1$, for $\alpha>1>\beta$ we
clearly have
\begin{equation}
\|q\|_{\alpha}\leq1\leq\|q\|_{\beta}
\, . \label{albe1}
\end{equation}
Hence, R\'{e}nyi entropies of discrete probability distributions
are always positive including zero for deterministic
distributions. This is not the case for differential entropies of
the form (\ref{redenw}). Despite of the normalization
$\|v\|_{1}=1$, we cannot generally write a continuous counterpart
of (\ref{albe1}). The quantity (\ref{redenw}) is
not of definite sign and becomes negative for density functions
with sufficiently large variations. Nevertheless, relations with
such entropies may express non-trivial conditions. For instance,
the differential Shannon entropy of phase with negative values was
considered in \cite{mjwh93}. Differential entropies are also an
intermediate point in obtaining conditions for entropies with
binning.

We will also use several mathematical
results for functions of one scalar variable. First, we recall the
Hausdorff--Young inequality with sharp constants. The question
concerns relations between norms of a function and its Fourier
transform. The sharp Hausdorff--Young inequality was found by
Beckner \cite{beckner} with using the previous result of Babenko
\cite{babenko}. We recall this result in a reformulation
convenient for our aims. It deals with probability density
functions and leads to entropic uncertainty relations for the
position and momentum \cite{birula1}. So, the Hausdorff--Young
inequality with sharp constants leads to an improvement of the
first entropic uncertainty relation of Hirschman \cite{hirs}. Let
two functions $\psi(x)$ and $\varphi(k)$ be connected by the Fourier
transform, namely
\begin{align}
\psi(x)&=\frac{1}{\sqrt{2\pi}}\int_{\cbr}
\exp(+\iu{k}x)\,\varphi(k)\,\xdif{k}
\, , \label{psxdf}\\
\varphi(k)&=\frac{1}{\sqrt{2\pi}}\int_{\cbr}
\exp(-\iu{k}x)\,\psi(x)\,\xdif{x}
\, . \label{phpdf}
\end{align}
They are treated as wave functions in the position and momentum
spaces, respectively. Here, we accepted units in which $\hbar=1$.
The probability density functions are written as
\begin{equation}
v(x)=|\psi(x)|^{2}
\, , \qquad
\tiv(k)=|\varphi(k)|^{2}
\, . \label{wxup}
\end{equation}
Introducing the Fourier transform by means of (\ref{phpdf}) is
physically motivated. At the same time, the formula (\ref{phpdf})
slightly differs from the definition commonly used in the mathematical
literature. So, we wrote the corresponding inequality of
Beckner \cite{beckner} in terms of the above wave functions and
converted it into relations between density
functions. We refrain from presenting the details here, since this
point has been addressed, e.g., in the works
\cite{IBB06,rastnum,barper15}. The result is posed as follows. Let
positive indices $\alpha$ and $\beta$ obey
$1/\alpha+1/\beta=2$, and let $\alpha>1>\beta$. For any quantum
state, norm-like functionals of the position and momentum
densities then obey
\begin{align}
\|v\|_{\alpha}
&\leq\left(\frac{1}{\varkappa\pi}\right)^{\!(1-\beta)/\beta}\|\tiv\|_{\beta}
\, , \label{wuxp}\\
\|\tiv\|_{\alpha}
&\leq\left(\frac{1}{\varkappa\pi}\right)^{\!(1-\beta)/\beta}\|v\|_{\beta}
\, . \label{uwpx}
\end{align}
Here, the positive parameter $\varkappa$ is given by the formula
\begin{equation}
\varkappa^{2}=\alpha^{1/(\alpha-1)}\beta^{1/(\beta-1)}
\, . \label{vkpf}
\end{equation}
It will be convenient to parametrize the indices $\alpha$
and $\beta$ as $1/\alpha=1-\tau$ and $1/\beta=1+\tau$ with
$\tau\in[0;1]$. By calculations, we then get
\begin{equation}
2\ln\varkappa(\tau)=\frac{1+\tau}{\tau}\,\ln(1+\tau)
-\frac{1-\tau}{\tau}\,\ln(1-\tau)
\, . \label{2vkap}
\end{equation}
The relations (\ref{wuxp}) and (\ref{uwpx}) hold, when wave
functions are related via the Fourier transform with infinite
limits. The corresponding observables obey the
position-momentum commutation relation and have eigenvalues
covering the real axis. Note that the sharp Hausdorff--Young
inequality {\it per se} implies (\ref{wuxp}) and (\ref{uwpx}) only
for pure states. However, they can immediately be extended to
mixed states. The ``twin'' relations (\ref{wuxp}) and (\ref{uwpx})
leads to uncertainty relations in terms of R\'{e}nyi's entropies
as described in \cite{IBB06,barper15}.

Dealing with convolutions, we should recall Young's inequality
with sharp constants. It was found by Beckner \cite{beckner} and
by Brascamp and Lieb \cite{bl76} with the use of different
methods. We will closely follow the formulation of Brascamp and
Lieb, since they also gave the converse of Young's inequality with
sharp constants. To each real index $a\geq1$, we assign the
conjugate index $a^{\prime}$ such that
\begin{equation}
\frac{1}{a}+\frac{1}{a^{\prime}}=1
\, . \label{nunup}
\end{equation}
The Young inequality involves factors of the form $C(a)$ defined
by
\begin{equation}
C(a)^{2}=a^{1/a}\,(a^{\prime})^{-1/a^{\prime}}
\, . \label{anuf}
\end{equation}
By $f*g$, we will mean the convolution of two functions of one
scalar variable. Let the indices be such that $a_{\ell},a\geq1$
and their conjugate ones obey
\begin{equation}
\sum_{\ell=1}^{n}\frac{1}{a_{\ell}^{\prime}}=\frac{1}{a^{\prime}}
\, . \label{aaj}
\end{equation}
For the convolution of $n$ functions, one has
\begin{equation}
C(a)\,\|f_{1}*\cdots*f_{n}\|_{a}
\leq\prod_{\ell=1}^{n}C(a_{\ell})\,\|f_{\ell}\|_{a_{\ell}}
\, . \label{yin1n}
\end{equation}
More results about the Young inequality as well as the
Hausdorff--Young inequality can be found, e.g., in chapters 4 and
5 of the book by Lieb and Loss \cite{lieb01}.

Inequalities converse to (\ref{yin1n}) generally
involve indices, some of which are negative. Here, the definition
should be reformulated. If indices $b$ and $b^{\prime}$ are
conjugate in the sense of (\ref{nunup}), then \cite{bl76}
\begin{equation}
C(b)^{2}=|b|^{1/b}\,|b^{\prime}|^{-1/b^{\prime}}
\, . \label{anuf1}
\end{equation}
When $0<b<1$, the conjugate index $b^{\prime}$ is strictly
negative. To emphasize this distinction, we prefer to mention
(\ref{anuf}) and (\ref{anuf1}) independently. Let the indices be
such that $0<b_{\ell},b\leq1$ and
\begin{equation}
\sum_{\ell=1}^{n}\frac{1}{b_{\ell}^{\prime}}=\frac{1}{b^{\prime}}
\, . \label{bbj}
\end{equation}
For the convolution of $n$ one-dimensional functions, we have
\cite{bl76}
\begin{equation}
\prod_{\ell=1}^{n}C(b_{\ell})\,\|f_{\ell}\|_{b_{\ell}}
\leq{C}(b)\,\|f_{1}*\cdots*f_{n}\|_{b}
\, . \label{cyin1n}
\end{equation}
This issue is connected with some previous results of Leindner and
Pr\'{e}kopa (see, e.g., references in \cite{bl76}). Calculating
with factors of the form (\ref{anuf}) and (\ref{anuf1}), we will
often use the expression
\begin{equation}
2\ln{C}(b)=\frac{1}{b^{\prime}}\,\ln\frac{1}{|b^{\prime}|}
-\frac{1}{b}\,\ln\frac{1}{|b|}
\, . \label{2cqp}
\end{equation}
In the following, both the inequalities (\ref{yin1n}) and
(\ref{cyin1n}) will be used in deriving separability conditions.
One form of the Minkowski inequality will also be recalled when
appropriate. Concerning this inequality, see corresponding
sections of the book by Hardy {\it et al.} \cite{hardy}.

\section{Entanglement criteria for a multipartite quantum system}\label{sec3}

In this section, we obtain $n$-separability conditions for an
$n$-partite quantum system with continuous variables. Let
subsystems of an $n$-partite system be labeled by
$\ell=1,\ldots,n$. The product
$\hh_{1:n}=\hh_{1}\otimes\cdots\otimes\hh_{n}$ is the total
Hilbert space. Any quantum state of the total system is given by a
density matrix $\bro_{1:n}$ on $\hh_{1:n}$. Density matrices are
assumed to be normalized. We note that $n$-fold product states of
the form $\bro_{1}\otimes\cdots\otimes\bro_{n}$ have no
correlations between subsystems. Recall that a bipartite mixed
state $\bro_{1:2}$ is called separable, when its density matrix
can be written as a convex combination of product states
\cite{werner89,zhsl98}. For an $n$-partite system, we call
$\bro_{1:n}$ to be $n$-separable, when it can be represented as a
convex combination of product states of the form
$\bro_{1}\otimes\cdots\otimes\bro_{n}$. Such states are often
called fully separable \cite{gt2009}. Without loss of generality,
each separable state will be treated as a convex combination of
only pure product states.

To formulate entanglement criteria, appropriate global observables
will be built from local ones \cite{duan00,mgvt02,gmvt03}. We
first recall the formulation for a bipartite system. To each
subsystem $\ell=1,2$, one assigns the position and momentum
variables $\bxx_{\ell}$ and $\bpx_{\ell}$ so that
$[\bxx_{\ell},\bpx_{\ell}]=\iu\pen_{\ell}$, where $\pen_{\ell}$ is
the identity on $\hh_{\ell}$. Using real $\theta_{\ell}$, we
define the operators
\begin{align}
\brx_{\ell}&:= \cos\theta_{\ell}\,\bxx_{\ell}+\sin\theta_{\ell}\,\bpx_{\ell}
\, , \label{brxj}\\
\bsx_{\ell}&:=-\sin\theta_{\ell}\,\bxx_{\ell}+\cos\theta_{\ell}\,\bpx_{\ell}
\, , \label{bsxj}
\end{align}
which also obey $[\brx_{\ell},\bsx_{\ell}]=\iu\pen_{\ell}$. It is
a linear canonical transformation in phase space, corresponding to
a unitary transformation of the Hilbert space \cite{wigner90}.

With the signs $\veps=\pm1$ and $\bar{\veps}=\mp1$, we further write
\begin{align}
\mrf_{\veps}&:=\brx_{1}\otimes\pen_{2}+\veps\,\pen_{1}\otimes\brx_{2}
\, , \label{grpm}\\
\msf_{\bar{\veps}}&:=\bsx_{1}\otimes\pen_{2}+\bar{\veps}\,\pen_{1}\otimes\bsx_{2}
\, . \label{gsmp}
\end{align}
The observables $\mrf_{\veps}$ and $\msf_{\bar{\veps}}$ are
commuting and jointly measurable. Let $|r_{\ell}\rangle$'s be
eigenkets of $\brx_{\ell}$ normalized through Dirac's delta
function. The observable (\ref{grpm}) satisfies
$\mrf_{\veps}|r_{1},r_{2}\rangle=(r_{1}+\veps\,r_{2})|r_{1},r_{2}\rangle$,
where $|r_{1},r_{2}\rangle=|r_{1}\rangle\otimes|r_{2}\rangle$. Let
$\bro_{1:2}$ be the state to be tested. For the observable
(\ref{grpm}), we get the probability density function after one
integration of
$\langle{r}_{1},r_{2}|\bro_{1:2}|r_{1},r_{2}\rangle$. For any
product state $\bro_{1}\otimes\bro_{2}$, we write
$w_{\ell}(r_{\ell})=\langle{r}_{\ell}|\bro_{\ell}|r_{\ell}\rangle$
so that this integrand reads as $w_{1}(r_{1})\,w_{2}(r_{2})$. The
probability density function of $r=r_{1}+\veps\,r_{2}$ then
becomes $w_{1}*w_{\veps2}$, where
$w_{\veps2}(r^{\prime})=w_{2}(\veps{r}^{\prime})$. Using this
fact, the authors of \cite{wtstd09} applied the entropy power
inequality for the Shannon entropy. For the R\'{e}nyi
$\alpha$-entropy, a version of entropy power inequalities was
given in \cite{bobch15} but only for $\alpha\geq1$. The problem of
extending the entropy power inequality to orders $0<\alpha<1$
remains open. To derive biseparability conditions in terms of
generalized entropies, the authors of \cite{stw11} used Young's
inequality and its converse with sharp constants. Entropy power
inequalities of the R\'{e}nyi type with improved coefficients
depend on number and dimensionality of involved random vectors
\cite{ram16}. As the dimensionality is explicitly used, such
inequalities are relevant only for finite-dimensional variables.

Global observables for an $n$-partite system with continuous
variables can be built in a similar way. To $\ell$-th subsystem,
we assign the operators $\widetilde{\brx}_{\ell}$ and
$\widetilde{\bsx}_{\ell}$, acting as $\brx_{\ell}$ and
$\bsx_{\ell}$ in $\hh_{\ell}$ and as the identity in other
subspaces. More precisely, we write
\begin{equation}
\widetilde{\brx}_{\ell}=
\pen_{1}\otimes\cdots\otimes\pen_{\ell-1}\otimes
\brx_{\ell}\otimes\pen_{\ell+1}\otimes\cdots\otimes\pen_{n}
\, , \label{wbrx}
\end{equation}
and similarly for $\widetilde{\bsx}_{\ell}$. Taking
$\veps_{\ell},\varep_{\ell}\in\{+1,-1\}$, the two observables of
interest are then defined as
\begin{align}
\mrf_{\veps:\veps}&:=
\veps_{1}\widetilde{\brx}_{1}+\cdots+\veps_{n}\widetilde{\brx}_{n}
\, , \label{ngrpm}\\
\msf_{\varep:\varep}&:=
\varep_{1}\widetilde{\bsx}_{1}+\cdots+\varep_{n}\widetilde{\bsx}_{n}
\, . \label{ngsmp}
\end{align}
The operators $\widetilde{\brx}_{\ell}$ and
$\widetilde{\bsx}_{\ell^{\prime}}$ commute for
$\ell\neq\ell^{\prime}$, whence
\begin{equation}
\bigl[\mrf_{\veps:\veps},\msf_{\varep:\varep}\bigr]
=\iu\pen_{1:n}\sum_{\ell=1}^{n}\veps_{\ell}\,\varep_{\ell}
\, , \label{comrs}
\end{equation}
where $\pen_{1:n}$ is the identity on $\hh_{1:n}$. For even $n$,
we can make (\ref{ngrpm}) and (\ref{ngsmp}) to be commuting. Say,
we setup $\varep_{\ell}=\veps_{\ell}$ for odd $\ell$ and
$\varep_{\ell}=-\,\veps_{\ell}$ for even $\ell$.

The product state $|r_{1},\ldots,r_{n}\rangle$ is an eigenstate of
(\ref{ngrpm}), corresponding to the eigenvalue
$r=\veps_{1}r_{1}+\cdots+\veps_{n}r_{n}$. Like the case $n=2$, we
define
\begin{equation}
V(r_{1},\ldots,r_{n})
=\langle{r}_{1},\ldots,r_{n}|\bro_{1:n}|r_{1},\ldots,r_{n}\rangle
\, , \label{rr1n}
\end{equation}
For brevity, we suppose $\veps_{\ell}=+1$ for all
$\ell=1,\ldots,n$. To the observable (\ref{ngrpm}), one assigns
the probability density function
\begin{equation}
W(r)=\underset{r_{1}+\cdots+r_{n}=r}{\,\int\cdots\int\,}
V(r_{1},\ldots,r_{n})\>\xdif{r}_{1}\cdots\xdif{r}_{n-1}
\, . \label{r1nn}
\end{equation}
For an $n$-fold product state
$\bro_{1}\otimes\cdots\otimes\bro_{n}$, the expression
(\ref{r1nn}) results in the convolution of $n$ local densities,
\begin{equation}
W=w_{1}*\cdots*w_{n}
\, . \label{nfcn}
\end{equation}
For other choices of the signs $\veps_{\ell}$, we replace each
$w_{\ell}$ with $w_{\veps\ell}$, where
$w_{\veps\ell}(r^{\prime})=w_{\ell}(\veps_{\ell}r^{\prime})$. The
same claims hold for probability density functions assigned to
(\ref{ngsmp}).

The above findings allow us to derive $n$-separability conditions in
terms of R\'{e}nyi entropies. We will follow the strategy already
justified in the previous papers
\cite{rastnum,barper15,rastcon16}. Here, the basic idea is to deal
with inequalities between norm-like functionals of the form
(\ref{alpnrm}). Desired relations in terms of suitable entropies
are extracted only at the final step. It is important that our
approach can naturally be combined with Young's inequality {\it
per se}. It is therefore more direct than appealing to entropy
power inequalities. Indeed, the entropy power inequality proved in
\cite{bobch15} are mainly based on Young's inequality with sharp
constants. The method of \cite{bobch15} was inspired by the
earlier Lieb's proof of the entropy power inequality for the
Shannon entropy \cite{lieb78}. The following statement takes place.

\newtheorem{prel}{Proposition}
\begin{prel}\label{pon31}
Let positive indices $a$ and $b$ be defined by the formulas
\begin{equation}
\frac{1}{a}=1-t
\, , \qquad
\frac{1}{b}=1+t
\, , \label{algam}
\end{equation}
where $t\in[0;1]$, and let
\begin{equation}
\ln\clk(t)=\frac{1}{2}\left(
\frac{1+t}{t}\,\ln(1+t)
-\frac{1-t}{t}\,\ln(1-t)
\right)
 . \label{clkt}
\end{equation}
Let $W$ and $U$ be density functions obtained respectively for the
observables (\ref{ngrpm}) and (\ref{ngsmp}) with $n\geq2$. If
$\,n$-partite state $\bro_{1:n}$ is $n$-separable, then we have the
inequality
\begin{equation}
H_{a}(W|\bro_{1:n})+H_{b}(U|\bro_{1:n})\geq\ln(n\clk\pi)
\, , \label{tw1n}
\end{equation}
and its ``twin'' with swapped $W$ and $U$.
\end{prel}

{\bf Proof.} To simplify the notation, we will take
$\veps_{\ell}=\varep_{\ell}=+1$ for all $\ell=1,\ldots,n$. For
other choices of the signs in (\ref{ngrpm}) and (\ref{ngsmp}), the
desired results follow due to
\begin{equation}
\|w_{\veps\ell}\|_{\alpha}=\|w_{\ell}\|_{\alpha}
\, , \qquad
\|u_{\varep\ell}\|_{\beta}=\|u_{\ell}\|_{\beta}
\, . \label{vevar}
\end{equation}
First, we will prove inequalities for arbitrary
product state. Writing $\tau=t/n\in[0;1/n]$, we further use
the parameters $\alpha$ and $\beta$ such that
\begin{equation}
\frac{1}{\alpha}=1-\tau
\, , \qquad
\frac{1}{\alpha^{\prime}}=\tau
\, , \qquad
\frac{1}{\beta}=1+\tau
\, , \qquad
\frac{1}{\beta^{\prime}}=-\,\tau
\, . \label{aabb}
\end{equation}
In Young's inequality (\ref{yin1n}), we set $a_{\ell}=\alpha\geq1$
for all $\ell=1,\ldots,n$, whence the index restriction
(\ref{aaj}) gives
\begin{equation}
\frac{1}{a^{\prime}}=\frac{n}{\alpha^{\prime}}=n\tau
\, , \qquad
\frac{1}{a}=1-n\tau
\, , \label{prest}
\end{equation}
consistently with (\ref{algam}) due to $t=n\tau$. Combining
(\ref{yin1n}) with (\ref{nfcn}) then gives
\begin{equation}
\|W\|_{a}\leq{C}(\alpha)^{n}\,C(a)^{-1}
\prod_{\ell=1}^{n}\|w_{\ell}\|_{\alpha}
\, . \label{form1}
\end{equation}
With each of the quantities $\|w_{\ell}\|_{\alpha}$, we use local
uncertainty relations of the form (\ref{wuxp}). This step results
in
\begin{equation}
\|W\|_{a}\leq\frac{C(\alpha)^{n}}{C(a)}\left(\frac{1}{\varkappa\pi}\right)^{\!{n}(1-\beta)/\beta}
\prod_{\ell=1}^{n}\|u_{\ell}\|_{\beta}
\, , \label{form2}
\end{equation}
where $\varkappa(\tau)$ is defined by (\ref{2vkap}). To the
convolution $U=u_{1}*\cdots*u_{n}$, we apply (\ref{cyin1n})
with setting $0<b_{\ell}=\beta\leq1$ for all $\ell=1,\ldots,n$.
Consistently with (\ref{algam}), the index restriction (\ref{bbj})
implies
\begin{equation}
\frac{1}{b^{\prime}}=\frac{n}{\beta^{\prime}}=-\,n\tau
\, , \qquad
\frac{1}{b}=1+n\tau
\, , \label{qrest}
\end{equation}
and the converse of Young's inequality reads
\begin{equation}
\prod_{\ell=1}^{n}\|u_{\ell}\|_{\beta}\leq
{C}(\beta)^{-n}\,C(b)\,\|U\|_{b}
\, . \label{form3}
\end{equation}
Combining (\ref{form2}) with (\ref{form3}) immediately gives
\begin{equation}
\|W\|_{a}\leq
\frac{C(\alpha)^{n}\,C(b)}{C(a)\,C(\beta)^{n}}\left(\frac{1}{\varkappa\pi}\right)^{\!t}
\|U\|_{b}
\, . \label{form4}
\end{equation}
It will be convenient to simplify factors that appeared in the
right-hand side of (\ref{form4}).

Using expressions of the form (\ref{2cqp}), we further obtain
\begin{align}
&\frac{1}{t}\,\ln\frac{C(\beta)^{n}}{C(\alpha)^{n}}
=\frac{1}{2\tau}
\,\Bigl(
-2\tau\ln\tau-(1+\tau)\ln(1+\tau)
\Bigr.
\nonumber\\
&\Bigl.
{}+(1-\tau)\ln(1-\tau)
\Bigr)
=-\ln\tau-\ln\varkappa
\, , \label{frab}
\end{align}
where $\ln\varkappa$ is written from (\ref{2vkap}). The expression
for $\ln\bigl(C(b)\big/C(a)\bigr)$ is obtained from
$\ln\bigl(C(\beta)\big/C(\alpha)\bigr)$ by replacing $\tau$ with
$t$, whence
\begin{equation}
\frac{1}{t}\,\ln\frac{C(a)}{C(b)}=\ln{t}+\ln\clk
\, , \label{frab1}
\end{equation}
Combining (\ref{frab}) and (\ref{frab1}) with (\ref{clkt}) finally
gives
\begin{equation}
\frac{1}{t}\,\ln\frac{C(a)\,C(\beta)^{n}}{C(\alpha)^{n}\,C(b)}+\ln\varkappa\pi=
\ln(n\clk\pi)
\, . \label{newf}
\end{equation}
Thus, we can finally rewrite (\ref{form4}) in the form
\begin{equation}
\|W\|_{a}\leq
\left(\frac{1}{n\clk\pi}\right)^{\!t}
\|U\|_{b}
\, . \label{form5}
\end{equation}
By a parallel argument, we can obtain the ``twin'' of (\ref{form5})
with swapped $W$ and $U$. The latter holds
for each product state. Before completing the proof, we should
extend our findings to separable states.

Each separable state can be represented as a convex combination of
product states. Hence, we obtain
\begin{equation}
W(r)=\sum\nolimits_{\lambda}\lambda\,W^{(\lambda)}(r)
\, , \label{wdenp}
\end{equation}
and a similar expression for $U$. Of course, the
weights are normalized here as $\sum_{\lambda}\lambda=1$.
Following \cite{IBB06,rast10r}, at this step we use the Minkowski
inequality \cite{hardy}. This inequality results in
\begin{align}
&\|W\|_{a}=\left\|\sum\nolimits_{\lambda}\lambda\,W^{(\lambda)}\right\|_{a}
\leq\sum\nolimits_{\lambda}\lambda\,\bigl\|W^{(\lambda)}\bigr\|_{a}
\, , \label{wmink}\\
&\sum\nolimits_{\lambda}\lambda\,\bigl\|U^{(\lambda)}\bigr\|_{b}
\leq\left\|\sum\nolimits_{\lambda}\lambda\,U^{(\lambda)}\right\|_{b}
=\|U\|_{b}
\, , \label{umink}
\end{align}
where we recall $a>1>b>0$. For each $\lambda$, the quantities
$\bigl\|W^{(\lambda)}\bigr\|_{a}$ and
$\bigl\|U^{(\lambda)}\bigr\|_{b}$ satisfy
(\ref{form5}). The latter remains therefore valid for the
quantities $\|W\|_{a}$ and $\|U\|_{b}$
calculated in any separable state.

The final step is to convert (\ref{form5}) into entropic
inequalities. We will first obtain entropic relations for $t>0$.
The Shannon case $a=b=1$ is reached by taking the corresponding
limit. The R\'{e}nyi entropies are expressed via norm-like
functionals according to (\ref{redenw1}), whence
\begin{align}
H_{a}(W|\bro_{1:n})&=-\frac{1}{t}\,\ln\|W\|_{a}
\, , \label{redenw21}\\
H_{b}(U|\bro_{1:n})&=\frac{1}{t}\,\ln\|U\|_{b}
\, . \label{redenw22}
\end{align}
To reach (\ref{tw1n}), we take the logarithm of both the sides of
(\ref{form5}) and use (\ref{redenw21}) and (\ref{redenw22}). The
inequality with swapped $W$ and $U$ is obtained by a very
parallel argument.
$\blacksquare$

We obtained $n$-separability conditions as a strictly positive
lower bound on the sum of two R\'{e}nyi entropies. With growth of
$n$, the lower bound in the right-hand side of (\ref{tw1n})
increases as a logarithm. For $n=2$, our result is similar to the
biseparability conditions derived in \cite{stw11}. The latter
generalizes the conditions in terms of differential Shannon
entropies proved in \cite{wtstd09}. Going from product states to
separable ones, the authors of \cite{wtstd09} used concavity of
the Shannon entropy. In general, a reference to concavity is not
relevant for R\'{e}nyi entropies. It is for this reason that the
Minkowski inequality was applied. In a similar manner, the
Minkowski inequality was already used in deriving entropic
uncertainty relations in \cite{IBB06,rast10r}. Here, we again see
a convenience of dealing  with relations between norm-like
functionals.

With growth of $t\in[0;1]$, the parameter (\ref{clkt}) decreases
from $\clk(0)=e$ up to $\clk(1)=2$. In terms of differential
Shannon entropies, we therefore have
\begin{equation}
H_{1}(W|\bro_{1:n})+H_{1}(U|\bro_{1:n})
\geq\ln({n}e\pi)
\, . \label{sepco0}
\end{equation}
For $n=2$, this inequality reduces to the main result of
\cite{wtstd09}. Thus, we have obtained an $n$-partite extension of
Shannon-entropy entanglement criteria for continuous variables. Of
course, concrete experimental setup is prescribed by the choice of
the angles $\theta_{\ell}$ in (\ref{brxj}) and (\ref{bsxj}). For the
input state of a $n$-partite system, we then measure commuting
observables (\ref{ngrpm}) and (\ref{ngsmp}). Evaluating the
densities $W$ and $U$, we can check the condition (\ref{tw1n}) and
its ``twin'' for various $t$. Their violation for any value of $t$
will imply that the input is not $n$-separable.

Using the entropy power inequality, the authors of \cite{wtstd09}
gave a combined inequality, which involves two global and four
local densities. Such relations merely reflect the fact that the
given density function is the convolution of two local densities.
To use them in entanglement detection, we must {\it a priori} be
sure that the input state is pure. This case will be exemplified
in Sect. \ref{sec5}. In general, however, the consideration of
only pure states is too idealized.

\section{Criteria in terms of discretized distributions}\label{sec4}

Previously, we have derived $n$-separability
conditions for continuous variables in terms of R\'{e}nyi
entropies. Such relations are not applicable
immediately in analysis of experimental data. Continuous-variable
probability density functions are typically replaced with
experimentally resolvable discrete probability distributions. In
the following, we aim to reformulate our separability conditions
due to the above reasons. In more detail, the problem of
entanglement detection under coarse-grained measurements was
examined in \cite{trgtw13}.

Entropic functions of the form (\ref{redenw}) may generally take
negative values. On the other hand, experiments typically result
in discrete probability distributions obtained by sampling density
functions of continuous variables. So, the density functions $W$
and $U$ will be used with a discretization into some bins. Let $W$
be sampled with respect to the set of prescribed marks
$\{\zeta_{j}\}$. Correspondingly, one puts the intervals
$\Delta\zeta_{j}=\zeta_{j+1}-\zeta_{j}$ and
$\Delta\zeta=\max\Delta\zeta_{j}$. The discrete distribution
$p_{\Delta\zeta}$ is formed by the probabilities
\begin{equation}
p_{j}:=\int\nolimits_{\zeta_{j}}^{\zeta_{j+1}}
W(r)\,\xdif{r}=\int_{\cbr} d_{j}^{(\zeta)}(r)\, W(r)\,\xdif{r}
\, . \label{pjdz}
\end{equation}
Here, $d_{j}^{(\zeta)}(r)$ is a boxcar function equal to $1$ for
$r$ between $\zeta_{j}$ and $\zeta_{j+1}$. The probabilities
(\ref{pjdz}) represent chances for the corresponding detection
positions \cite{trgtw13}. The quantity $\Delta\zeta$ characterizes
the width of detectors in $r$-space. Similarly, the distribution
$q_{\Delta\xi}$ is gained by sampling $U$ with respect to bins
between marks $\xi_{i}$ so that $\Delta\xi_{i}=\xi_{i+1}-\xi_{i}$
and $\Delta\xi=\max\Delta\xi_{i}$.

There are two ways to express results of measurements of the
discussed type \cite{trgtw13}. First, we can deal immediately with
the discrete distributions $p_{\Delta\zeta}$ and $q_{\Delta\xi}$.
Second, we can construct approximations to the original
probability distributions $W(r)$ and $U(s)$, namely
\begin{align}
W_{\Delta\zeta}(r)&:=\sum_{j=-\infty}^{+\infty}
\Delta\zeta_{j}^{-1}p_{j}\,d_{j}^{(\zeta)}(r)
\, , \label{wapp}\\
U_{\Delta\xi}(s)&:=\sum_{i=-\infty}^{+\infty}
\Delta\xi_{i}^{-1}q_{i}\,d_{i}^{(\xi)}(s)
\, . \label{uapp}
\end{align}
The original probability densities are replaced with approximate
continuous distribution in the histogram form. When the bins all
tend to zero, these histograms reproduce the original
distributions.

Assuming $a>1>b>0$, we can prove the inequalities \cite{barper15,rastcon16}
\begin{align}
\Delta\zeta_{j}^{1-a}\,p_{j}^{a}&\leq
\int\nolimits_{\zeta_{j}}^{\zeta_{j+1}}W(r)^{a}\,\xdif{r}
\, , \label{probwu1}\\
\int\nolimits_{\xi_{i}}^{\xi_{i+1}}U(s)^{b}\,\xdif{s}
&\leq\Delta\xi_{i}^{1-b}\,q_{i}^{b}
\, . \label{probwu2}
\end{align}
These formulas are based on theorem 192 of \cite{hardy}. Another
way refers to an integral analog of Jensen's inequality with
weight functions of the form $d_{j}^{(\zeta)}(r)/\Delta\zeta_{j}$
\cite{rastmin16}. It follows from (\ref{wapp}) and (\ref{uapp})
that
\begin{align}
\bigl\|W_{\Delta\zeta}\bigr\|_{a}^{a}&=
\sum_{j=-\infty}^{+\infty}
\Delta\zeta_{j}^{1-a}\,p_{j}^{a}
\, , \label{nnwu1}\\
\bigl\|U_{\Delta\xi}\bigr\|_{b}^{b}&=
\sum_{i=-\infty}^{+\infty}
\Delta\xi_{i}^{1-b}\,q_{i}^{b}
\, . \label{nnwu2}
\end{align}
For $a>1>b$, we therefore obtain
\begin{align}
&\Delta\zeta^{1-a}\,\bigl\|p_{\Delta\zeta}\bigr\|_{a}^{a}
\leq\bigl\|W_{\Delta\zeta}\bigr\|_{a}^{a}\leq\|W\|_{a}^{a}
\, , \label{conv1}\\
&\|U\|_{b}^{b}\leq\bigl\|U_{\Delta\xi}\bigr\|_{b}^{b}
\leq\Delta\xi^{1-b}\,\bigl\|q_{\Delta\xi}\bigr\|_{b}^{b}
\, . \label{conv2}
\end{align}
These inequalities follow from combining (\ref{probwu1}) with
(\ref{nnwu1}) and (\ref{probwu2}) with (\ref{nnwu2}). Note also
that R\'{e}nyi's entropies of discrete distributions become
unbounded, when the size of bins tends to zero. It can be observed
from (\ref{conv1}) and (\ref{conv2}). In this limit, we will use
entropic separability conditions of the form (\ref{tw1n}).

Combining (\ref{form5}) with (\ref{conv1}) and (\ref{conv2}), the
following conclusions take place. For $n$-separable states of an
$n$-partite system, the result (\ref{form5}) and its ``twin''
remain valid for the histogram functions (\ref{wapp}) and
(\ref{uapp}). Further, we get the inequality
\begin{equation}
\bigl\|p_{\Delta\zeta}\bigr\|_{a}\leq
\left(\frac{\Delta\zeta\Delta\xi}{n\clk\pi}\right)^{\!t}
\bigl\|q_{\Delta\xi}\bigr\|_{b}
\, , \label{mrof5pq}
\end{equation}
and its ``twin'' with swapped $p_{\Delta\zeta}$ and $q_{\Delta\xi}$.
Here, the indices $a$ and $b$ are again defined by
(\ref{algam}). Entropic separability conditions with binning are
derived from (\ref{mrof5pq}) similarly to the way by which the
result (\ref{tw1n}) follows from (\ref{form5}).

\newtheorem{pqbin}[prel]{Proposition}
\begin{pqbin}\label{pon42}
Let positive indices $a$ and $b$ be defined for $t\in[0;1]$ by
(\ref{algam}), and let $\clk(t)$ be defined by (\ref{clkt}). Let
$p_{\Delta\zeta}$ and $q_{\Delta\xi}$ be distributions obtained
respectively by sampling the density functions in measurements of
(\ref{ngrpm}) and (\ref{ngsmp}) with $n\geq2$. Let
$W_{\Delta\zeta}$ and $U_{\Delta\xi}$ be the corresponding
histogram functions. If $n$-partite state $\bro_{1:n}$ is
$n$-separable, then we have the inequalities
\begin{align}
H_{a}\bigl(W_{\Delta\zeta}\big|\bro_{1:n}\bigr)+
H_{b}\bigl(U_{\Delta\xi}\big|\bro_{1:n}\bigr)
&\geq\ln(n\clk\pi)
\, , \label{tw1app}\\
H_{a}\bigl(p_{\Delta\zeta}\big|\bro_{1:n}\bigr)+
H_{b}\bigl(q_{\Delta\xi}\big|\bro_{1:n}\bigr)
&\geq\ln\!\left(\frac{n\clk\pi}{\Delta\zeta\Delta\xi}\right)
 , \label{tw1bin}
\end{align}
and their ``twins'' with swapped histogram functions and
probability distributions.
\end{pqbin}

The above entropic $n$-separability conditions are formulated
using distributions with discretization. The condition
(\ref{tw1bin}) extends R\'{e}nyi-entropy biseparability conditions
derived in \cite{stw11}. As experimentally resolvable
distributions are typically discrete, relations with such
distributions are more appropriate in practice. In the case $t=0$,
the formula (\ref{tw1app}) reads as
\begin{equation}
H_{1}\bigl(W_{\Delta\zeta}\big|\bro_{1:n}\bigr)+
H_{1}\bigl(U_{\Delta\xi}\big|\bro_{1:n}\bigr)
\geq\ln({n}e\pi)
 . \label{tw1bit0}
\end{equation}

We derived a one-parameter family of $n$-separability conditions
for an $n$-partite continuous-variable system. A utility of
entropic expressions with freely-variable parameters was noted in
\cite{maass} with respect to uncertainty relations. A family of
relations is more informative in the sense that it generally
provides stronger restrictions on involved probabilities. In
entanglement detection, a dependence on entropic parameter can be
used in slightly another manner. Varying the control parameter
$t$, we try to observe the violation of separability conditions.
When the violation has happened, we do detect entanglement of the
state to be tested. Of course, the violation of separability
conditions is sufficient but not necessary. There exist entangled
states that will escape the entanglement detection by particular
criteria. Nevertheless, a range of detectability will generally
increase with adding more separability conditions.

Note that the inequality (\ref{mrof5pq}) also leads to
separability conditions in terms of Tsallis entropies. Such
conditions can be obtained due to the minimization task of
\cite{rast104}. We present only the final result, since the
derivation {\it per se} was in detail considered in
\cite{barper15,rastcon16}. For $0<\alpha\neq1$, the Tsallis
$\alpha$-entropy of distribution $\{q_{i}\}$ is defined as
\cite{tsallis}
\begin{equation}
T_{\alpha}(q):=\frac{1}{1-\alpha}\left(\sum\nolimits_{i} q_{i}^{\alpha}-1\right)
=-\sum\nolimits_{i} q_{i}^{\alpha}\ln_{\alpha}(q_{i})
\, . \label{rsdf}
\end{equation}
For brevity, we use here the $\alpha$-logarithm
$\ln_{\alpha}(x)=\bigl(x^{1-\alpha}-1\bigl)/(1-\alpha)$. Let
positive indices $a$ and $b$ be defined for $t\in[0;1]$ by
(\ref{algam}), and let $\clk(t)$ be defined by (\ref{clkt}). If
$n$-partite state $\bro_{1:n}$ is $n$-separable, then we have the
inequality
\begin{equation}
T_{a}\bigl(p_{\Delta\zeta}\big|\bro_{1:n}\bigr)+
T_{b}\bigl(q_{\Delta\xi}\big|\bro_{1:n}\bigr)
\geq\ln_{a}\!\left(\frac{n\clk\pi}{\Delta\zeta\Delta\xi}\right)
 , \label{ts1bin}
\end{equation}
and its ``twin'' with swapped $p_{\Delta\zeta}$ and
$q_{\Delta\xi}$. Tsallis entropies of continuously changed
variables were also considered in the literature. However, the
minimization problem of \cite{rast104} is not applicable for
differential entropies.

Using entropies of discrete probability distributions, we can take
into account possible inefficiencies of the detectors used. In
practice, measurement devices inevitably suffer from losses.
Hence, some discussion of cases with non-zero probability of the
no-click event is of interest. The following simple model will be
considered. Let the parameter $\eta\in[0;1]$ characterize a
detector efficiency. To the given value $\eta$ and probability
distribution $\{q_{i}\}$, we assign a ``distorted'' distribution
such that
\begin{equation}
q_{i}^{(\eta)}=\eta{q}_{i}
\ , \qquad
q_{\varnothing}^{(\eta)}=1-\eta
\ . \label{petad}
\end{equation}
Here, the probability $q_{\varnothing}^{(\eta)}$ corresponds to
the no-click event. We further assume that in both the
measurements the inefficiency-free distributions are altered
according to (\ref{petad}). So, an efficiency of detection is
taken to be equal for all bins. The above
formulation is inspired by the first model of detection
inefficiencies used by the authors of \cite{rchtf12} in
the context of cycle scenarios of the Bell type. Since the
separability conditions (\ref{tw1bin}) and (\ref{ts1bin}) involve
different entropic parameters, we restrict a consideration to the
Shannon entropies. It is easy to check that
\begin{equation}
H_{1}\bigl(q^{(\eta)}\bigr)=\eta{H}_{1}(q)+h_{1}(\eta)
\, , \label{adbin}
\end{equation}
where $h_{1}(\eta)=\!{}-\eta\ln\eta-(1-\eta)\ln(1-\eta)$ is the
binary Shannon entropy . If $n$-partite state $\bro_{1:n}$ is
$n$-separable, then
\begin{equation}
H_{1}\bigl(p_{\Delta\zeta}^{(\eta)}\big|\bro_{1:n}\bigr)+
H_{1}\bigl(q_{\Delta\xi}^{(\eta)}\big|\bro_{1:n}\bigr)
\geq\eta\ln\!\left(\frac{{n}e\pi}{\Delta\zeta\Delta\xi}\right)+2h_{1}(\eta)
\, . \label{h1bineta}
\end{equation}
Thus, detector inefficiencies will produce additional
uncertainties in the entropies of actually measured data. With
decreasing $\eta>1/2$, the first term in the right-hand side of
(\ref{h1bineta}) reduces, whereas the second term increases. When
$\eta$ does not approach $1$ sufficiently closely, this feature
will prevent a robust detection of entanglement.

\section{Examples of application of the derived criteria}\label{sec5}

Finally, we shall illustrate a relevance of the presented
entanglement criteria with entropic parameters. The aim is not a
study of numerous types of continuous-variable states in full
detail. We rather wish to exemplify principal features of the new
$n$-separability conditions. Nevertheless, considered states may
be of interest in practice.

Our first example concerns $n$-partite states that are similar to
dephased cat states. By $|z\rangle$, we mean the coherent state
corresponding to complex number $z$. For $0\leq{c}\leq1$, we
define
\begin{align}
&\vbro_{1:n}(z)=
\cln(z)\,\Bigl\{
|z^{\otimes{n}}\rangle\langle{z}^{\otimes{n}}|+
|(-z)^{\otimes{n}}\rangle\langle(-z)^{\otimes{n}}|
\Bigr.
\nonumber\\
\Bigl.
&{}-(1-c)\bigl[
|z^{\otimes{n}}\rangle\langle(-z)^{\otimes{n}}|+
|(-z)^{\otimes{n}}\rangle\langle{z}^{\otimes{n}}|
\bigr]
\Bigr\}
\, , \label{decsf}
\end{align}
where $|z^{\otimes{n}}\rangle$ denotes the $n$-fold product state
and $\cln(z)$ is the normalization factor. For $n=2$, the formula
(\ref{decsf}) gives a bipartite dephased cat state. Such states
were used in order to test entanglement criteria in terms of
Shannon entropies \cite{wtstd09} and R\'{e}nyi entropies
\cite{stw11}. Applications of cat states in quantum information
processing with continuous variables are reviewed in
\cite{jeong07}.

To relate with the results of \cite{wtstd09,stw11}, we substitute
$\theta_{\ell}=0$ in (\ref{brxj}) and (\ref{bsxj}). As was already
mentioned, for even $n$ the observables (\ref{ngrpm}) and
(\ref{ngsmp}) can be made commuting. For $n=2m\geq2$, we take
the observables
\begin{align}
\mrf_{\veps:\veps}&=
\sum_{\ell=1}^{n}(-1)^{\ell-1}\,\widetilde{\brx}_{\ell}
\, , \label{mrfct}\\
\msf_{\varep:\varep}&=
\sum_{\ell=1}^{n}\,\widetilde{\bsx}_{\ell}
\, . \label{msfct}
\end{align}
When $n=2$, the formulas (\ref{mrfct}) and (\ref{msfct})
respectively lead to (\ref{grpm}) and (\ref{gsmp}) with $\veps=-1$
and $\bar{\veps}=+1$. Namely these commuting observables were used
in \cite{wtstd09,stw11}. Further, the family of states
(\ref{decsf}) will be considered for even $n$ and positive real
$z$. To study the violation of separability conditions, we
introduce the characteristic quantity
\begin{equation}
Q_{a}(z):=
\ln(n\clk\pi)-H_{a}(U|\vbro_{1:n})-H_{b}(W|\vbro_{1:n})
\, . \label{caz1n}
\end{equation}
Here, the indices $a$ and $b$ are linked by (\ref{algam}) and
$\clk(t)$ is defined by (\ref{clkt}). Strictly positive values of
$Q_{a}(z)$ will show that the tested state is entangled.

\begin{figure}
\includegraphics[width=7.6cm]{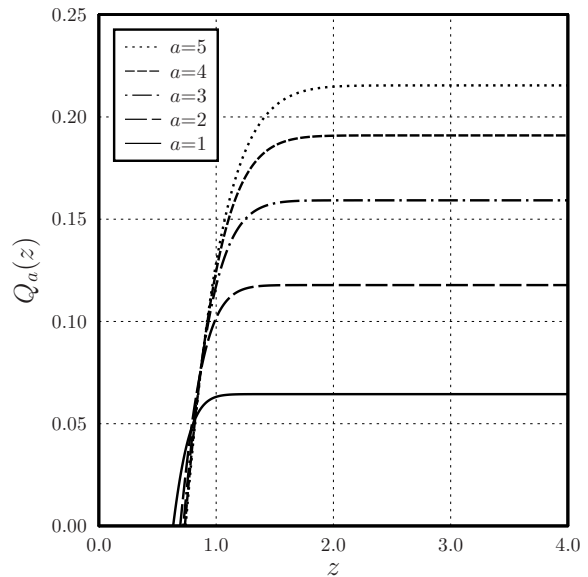}
\caption{\label{fg1} $Q_{a}(z)$ as a function of positive real $z$
for $n=4$, $c=1/2$, and $a=1,2,3,4,5$.}
\end{figure}

The characteristic quantity $Q_{a}(z)$ is drawn in Fig. \ref{fg1}
for $n=4$, $c=1/2$, and five values of $a$. Only positive values
are shown here. We also restrict a consideration to values
$z\in[0;4]$, since an asymptotic behavior already appears on the
right sides of curves. For a very large range of $z$, the
separability conditions in terms of R\'{e}nyi entropies allow to
detect entanglement. We also see that the undetectable region is
almost the same for all curves. Although the border of detectable
values becomes leftmost for the standard case $a=1$, this
difference is quite small and hardly significant in practice.
Indeed, all real devices are inevitably exposed to noise. In
opposite, the size of violation essentially depends on entropic
indices. With growth of $a$, the size of violation for
sufficiently large $z$ is increased more than three times. Thus,
separability conditions in terms of generalized entropies lead to
more robust detection of entanglement.

\begin{figure}
\includegraphics[width=7.6cm]{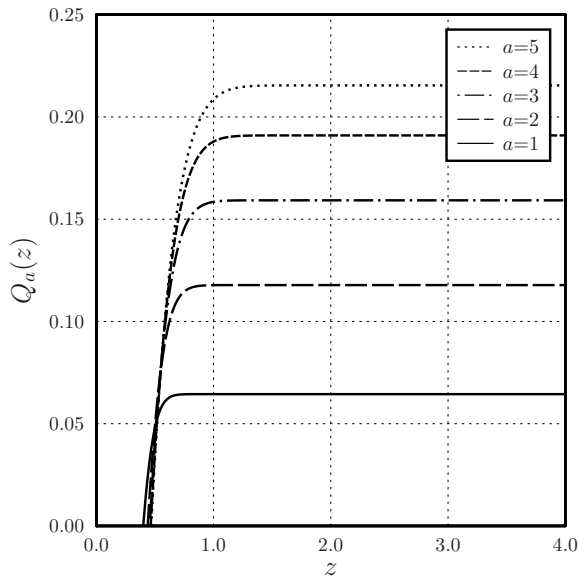}
\caption{\label{fg2} $Q_{a}(z)$ as a function of positive real $z$
for $n=10$, $c=1/2$, and $a=1,2,3,4,5$.}
\end{figure}

The obtained entanglement criteria have also been tested for even
$n>4$. A behavior of the curves is very similar to what we saw in
Fig. \ref{fg1}. With growing $a$ in the range considered, the size
of violation for sufficiently large $z$ is increased. Also, the
curves drawn for different $a$ have almost the same undetectable
region. So, we again see a significance of entanglement criteria
in terms of R\'{e}nyi entropies. On the other hand, the border of
detectable values goes to the left with growth of $n$. Increasing
$n$, the undetectable region of entanglement criteria becomes more
and more narrow. For example, we present $Q_{a}(z)$ in Fig.
\ref{fg2} for $n=10$, $c=1/2$, and five values of $a$. Comparing
Figs. \ref{fg1} and \ref{fg2}, one sees that the curves with
growth of $n$ try to approach a form like Heaviside's step
function.

The curves were presented for $c=1/2$, but for other values
$c\neq1$ we have seen a similar picture. Using generalized
entropies generally increases the size of violation. On the other
hand, growing $c$ implies a decrease of the factor $1-c$. Here,
the curves reveal the following tendencies. When other parameters
are fixed, positive values of $Q_{a}(z)$ are visibly reduced in
size. Furthermore, the undetectable region on the $z$-axis is
widen due to increasing $c$. The character of dependence on $c$
allows one to explain some natural relation between detectability
of entanglement for different values of $n$. Taking the partial
trace, we will obtain states of the same type (\ref{decsf}) but
with lesser $n$. When other parameters are fixed, the size of
violation increases and the undetectable region narrows with
growth of $n$. Another point is that the term $1-c$ will be
decreased due to tracing out some particles. Chances to detect
entanglement of states of the considered type cannot be improved
by applying the above entropic criteria to partial traces.

The size of violation of separability conditions is significant
due to the following reasons. In practice, the original
density functions $U$ and $W$ can additionally be masked in
experiments due to a finiteness of resolution and external noise.
These features can only increase the amount of uncertainty.
Instead of the theoretical violation (\ref{caz1n}), measurements
result in another quantity $\widetilde{Q}_{a}(z)$ such that
$\widetilde{Q}_{a}(z)\leq{Q}_{a}(z)$. Since actually observed
violation is reduced, our possibilities to detect entanglement
essentially depend on the size of violation of separability
conditions. To reach robust detection of entanglement, one will
try to maximize the characteristic quantity with respect to
entropic indices. So, the presented separability conditions are of
interest in practice of quantum information processing. In this regard, 
the new criteria provided an extension of basic results of \cite{stw11} 
to the case of multipartite systems.

\begin{figure}
\includegraphics[width=7.6cm]{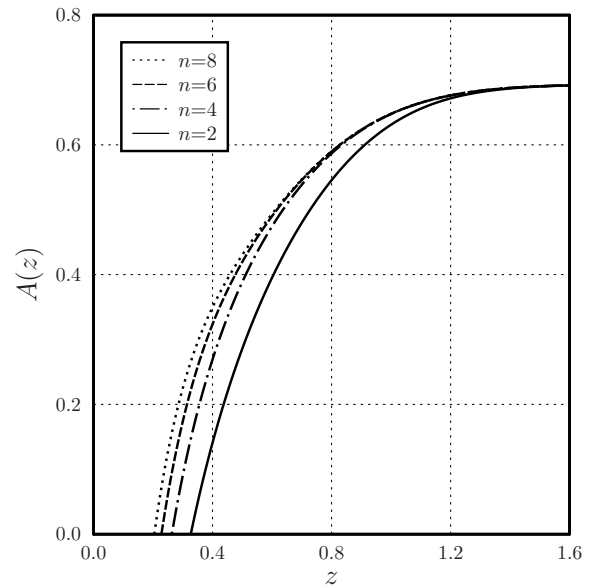}
\caption{\label{fg3} $A(z)$ as a function of positive real $z$
for $n=2,4,6,8$.}
\end{figure}

Let us consider briefly applications of the derived entropic
criteria to pure states. In this case, we can use the inequalities
(\ref{form1}) and (\ref{form3}) separately. Separable pure states
are written in the form
\begin{equation}
|\varPhi_{1:n}\rangle=\bigotimes_{\ell=1}^{n}|\phi_{\ell}\rangle
\, . \label{phipu}
\end{equation}
For such states, the condition (\ref{form1}) can be reformulated as
\begin{equation}
H_{a}(W|\,\varPhi_{1:n})\geq
\frac{1}{t}\,\ln\frac{C(a)}{C(\alpha)^{n}}
+\frac{1}{n}
\sum_{\ell=1}^{n}H_{\alpha}(u_{\ell}|\phi_{\ell})
\, , \label{form1pu}
\end{equation}
where the indices $a\geq1$ and $\alpha\geq1$ are such that their
conjugate ones obey (\ref{prest}). Assuming $n=2m\geq2$, we shall
test the criterion (\ref{form1pu}) with states of the form
\begin{align}
|\varPsi_{1:n}(z)\rangle=\sqrt{\clom(z)}
\>\Bigl\{
&|z^{\otimes{m}}\rangle\otimes|(-z)^{\otimes{m}}\rangle
\Bigr.
\nonumber\\
&\Bigl.
{}-|(-z)^{\otimes{m}}\rangle\otimes|z^{\otimes{m}}\rangle
\Bigr\}
 , \label{vpsin}
\end{align}
where $z\neq0$ and $\clom(z)$ is the normalization factor. We also
rewrite the sum (\ref{mrfct}) with the same sign for all summands.
Numerical calculations showed that we cannot reach a valuable
increase of violation of (\ref{form1pu}) by varying entropic
parameters. This situation is opposite to the curves drawn in
Figs. \ref{fg1} and \ref{fg2}. When the separability condition
(\ref{form1pu}) is applied to states of the form (\ref{vpsin}), a
picture somehow depends on the number of subsystems. Due to these
reasons, we further take $t=\tau=0$ and $a=\alpha=1$. On the other
hand, curves for different values of $n$ will be compared.
Similarly to (\ref{caz1n}), we put the characteristic quantity
\begin{equation}
A(z):=\frac{\ln{n}}{2}+
\frac{1}{n}\sum_{\ell=1}^{n}H_{\alpha}(w_{\ell}|\bro_{\ell})-H_{1}(W|\varPsi_{1:n})
\, , \label{zac1n}
\end{equation}
where $\bro_{\ell}$ is the corresponding partial trace. In Fig.
\ref{fg3}, we present positive values of $A(z)$ for $n=2,4,6,8$
and positive real $z$. For sufficiently large $z$, all the curves
approach the limiting value $\ln2\approx0.693$. So, we restrict a
consideration to values $z\in[0;1.6]$. Like the above examples,
the border of detectable values goes to the left with growth of
$n$. At the same time, a distinction between the curves is not so
essential as in Figs. \ref{fg1} and \ref{fg2}. Except for a small
region, we  have seen a violation of the $n$-separability
condition (\ref{form1pu}) up to arbitrary positive $z$.

Finally, we shall discuss possible applications of the presented
$n$-separability conditions in multipartite entanglement
detection. Our entanglement criteria are expressed in terms of
experimentally measured quantities with the use of sufficiently
simple and universal setup. In principle, states of an $n$-partite
system may be $n^{\prime}$-separable with $1<n^{\prime}\leq{n}$.
The equality $n^{\prime}=n$ implies that the given state is fully
separable. Our separability conditions allow one to test full
separability immediately. Their violation is sufficient for the
conclusion that the tested state is not fully separable. They may
also be used in a more complicated manner, when corresponding
partial traces of the input $\bro_{1:n}$ will be tested with
respect to full separability. This complexity is natural, since
the separability problem increases substantially in the context of
multipartite systems \cite{gt2009,sv13,gsv15,gsv16}.

\acknowledgments

I am grateful to Y.~Huang and W.~Vogel for useful feedback.

\end{document}